\documentclass[twocolumn,prl,aps,superscriptaddress]{revtex4-1}
\usepackage[latin9]{inputenc}
\usepackage{mathtools}
\usepackage{amsbsy}
\usepackage{amstext}
\usepackage[unicode=true,pdfusetitle,
 bookmarks=false,
 breaklinks=false,pdfborder={0 0 1},backref=false,colorlinks=false]
 {hyperref}
\hypersetup{
 bookmarksnumbered=false,bookmarksopen=false}
\makeatletter

\@ifundefined{textcolor}{}
{%
 \definecolor{BLACK}{gray}{0}
 \definecolor{WHITE}{gray}{1}
 \definecolor{RED}{rgb}{1,0,0}
 \definecolor{GREEN}{rgb}{0,1,0}
 \definecolor{BLUE}{rgb}{0,0,1}
 \definecolor{CYAN}{cmyk}{1,0,0,0}
 \definecolor{MAGENTA}{cmyk}{0,1,0,0}
 \definecolor{YELLOW}{cmyk}{0,0,1,0}
}

\usepackage{amsmath}
\usepackage{graphicx}
\usepackage{amssymb}
\usepackage{txfonts,color}
\makeatother

\begin{document}
\title{Phase-adaptive dynamical decoupling methods for robust spin-spin dynamics in trapped ions}

\author{Lijuan Dong}
\affiliation{International Center of Quantum Artificial Intelligence for Science and Technology~(QuArtist)
and Department of Physics, Shanghai University, 200444 Shanghai, China}
\affiliation{Department of Physical Chemistry, University of the Basque Country UPV/EHU, Apartado 644, 48080 Bilbao, Spain}

\author{I\~nigo Arrazola}
\affiliation{Department of Physical Chemistry, University of the Basque Country UPV/EHU, Apartado 644, 48080 Bilbao, Spain}
\author{Xi Chen}
\affiliation{International Center of Quantum Artificial Intelligence for Science and Technology~(QuArtist)
and Department of Physics, Shanghai University, 200444 Shanghai, China}
\affiliation{Department of Physical Chemistry, University of the Basque Country UPV/EHU, Apartado 644, 48080 Bilbao, Spain}
\author{Jorge Casanova}
\affiliation{Department of Physical Chemistry, University of the Basque Country UPV/EHU, Apartado 644, 48080 Bilbao, Spain}
\affiliation{IKERBASQUE,  Basque  Foundation  for  Science, Plaza Euskadi 5, 48009 Bilbao,  Spain}

\begin{abstract}
Quantum platforms based on trapped ions are main candidates to build a quantum hardware with computational capacities that largely surpass those of classical devices. Among the available control techniques in these setups, pulsed dynamical decoupling (pulsed DD) revealed as a useful method to process the information encoded in ion registers, whilst minimising the environmental noise over them. In this work, we incorporate a pulsed DD technique that uses random pulse phases, or correlated pulse phases, to significantly enhance the robustness of entangling spin-spin dynamics in trapped ions. This procedure was originally conceived in the context of nuclear magnetic resonance for nuclear spin detection purposes, and here we demonstrate that the same principles apply for robust quantum information processing in trapped-ion settings.
\end{abstract}
\maketitle

\section{Introduction}
The controlled generation of arbitrary quantum states, their subsequent processing via quantum operations, and their adequate preservation in quantum bits (qubits) are recurrent duties in the quantum computing area~\cite{Nielsen00,Dowling03}. In platforms such as superconducting circuits and trapped ions~\cite{Haffner08,Blais20}, bosonic modes that couple several qubits enables to construct  arbitrary quantum states via entangling gates among several qubits. In particular, in setups based on trapped ions, qubits are encoded in the internal degrees of freedom of the ions, whilst collective vibrational normal modes of the ion chain are used as a bosonic quantum bus. This mediation, along with the ability to control the internal states of the atoms with laser and/or microwave (MW) radiation~\cite{Leibfried03,Bardin20}, led to high-fidelity quantum information processing in the form of entangling gates~\cite{Ballance16,Gaebler16}, or quantum simulations of many-body spin systems~\cite{Friedenauer08,Kim10,Bermudez11,Britton12,Islam13,Jurcevic14,Richerme14}. 

In trapped ions, qubit-boson coupling is typically achieved with a laser that supplies the energy gap between the ion qubit and  specific vibrational modes~\cite{Leibfried03}. Here, the strength of the qubit-boson interaction is proportional to the Lamb-Dicke (LD) factor $\eta$. This is given by the ratio between the width of the motional ground-state wavefunction, and the wavelength of the employed laser. Often, the LD factor takes values below $0.1$~\cite{Leibfried03}, such as $\eta=0.044$ in Ref.~\cite{Gerritsma11} or $\eta=0.089$ in Ref.~\cite{Um16}. On the other hand, one can also use trapped ions that encode a qubit with an energy splitting lying on the MW regime. One example of this is the $^{171}$Yb$^{+}$ ion~\cite{Balzer06,Olmschenk07}. In this scenario, MW fields can exert local operations on the qubit states, but fail to induce a coupling with the vibrational modes of the ion chain. This owes to the effective LD factor of the qubit-boson interaction that amounts to $\sim10^{-7}$ as a consequence of the long wavelength of the MW radiation fields. Solutions to this problem involve introducing either static~\cite{Mintert01,Welzel11,Khromova12,Piltz16,Wolk17,Welzel19} or oscillating~\cite{Ospelkaus08,Ospelkaus11,Hanh19,Zarantonello19,Srinivas19} magnetic-field gradients. In current setups, these gradients induce an effective LD factor which is approximately one order of magnitude smaller than those in laser-based systems. For example, the MW scheme in Ref.~\cite{Weidt16} leads to a LD factor $\eta=0.0041$. However, larger values for the LD factor are expected to be reached in new MW setups~\cite{Weidt16}. From a different perspective, it is noteworthy  to mention that the use of MW radiation has several advantages with respect to lasers. On the one hand, MW technology is easier to control and incorporate in scalable trap designs~\cite{Lekitsch17}. On the other hand, MW radiation avoids spontaneous scattering of photons, which is a fundamental limitation in laser-based ion setups~\cite{Ballance16,Gaebler16}. Furthermore, long-wavelength radiation is less sensitive to phase perturbations due to, for example, mirror vibrations or geometrical drifts.

In another vein, qubit-boson coupling can be achieved in two different fashions. On the one hand, a transversal coupling of the form $H_x=\eta\Omega(a+a^\dagger)\sigma_x$ is achieved in laser-driven interactions~\cite{Roos08} or using magnetic-field gradients~\cite{Sutherland19}. On the other hand, longitudinal couplings of the form $H_z=\eta\Omega(a+a^\dagger)\sigma_z$ are produced  with lasers~\cite{Britton12,Roos08,Belmechri13}, as well as with static~\cite{Mintert01} or oscillating~\cite{Sutherland19} magnetic-field gradients. An important difference between transversal and longitudinal couplings is that the latter commutes with the system's bare Hamiltonian. This is defined by the $\sigma_z|e\rangle=|e\rangle$ and $\sigma_z|g\rangle=-|g\rangle$ basis states. As a consequence, longitudinal couplings enable a straightforward application of pulsed DD techniques~\cite{Hahn50,Carr54,Meiboom58} while, typically, continuous DD schemes are pursued for transversal couplings~\cite{Bermudez12,Lemmer13,Cohen15,Puebla16,Arrazola20}. Both techniques, pulsed and continuous, are  devoted to cancel errors leading to qubit dephasing. However, pulsed DD schemes offer a superior performance against environmental fluctuations, as well as in front of errors on the drivings.

Besides protecting the system from environmental noise, DD techniques are routinely employed to exert  control on  the system evolution. For example, a qubit can be dynamically decoupled from environmental signals excepting from those whose frequencies lie on a certain energy window. This enables to couple a qubit with specific signals leading to  applications in quantum sensing~\cite{Alvarez11,BarGill12,Casanova15,Baumgart16,Arrazola19,Munuera20}. In trapped ions, DD has been used to accomplish high-fidelity entangling gates~\cite{Weidt16,Harty16}. Furthermore, pulsed DD has been proposed as generator of fast gates with MW-driven ions~\cite{Arrazola18}.

In this article we demonstrate that phase-adaptive DD methods, previously introduced in the context of nuclear magnetic resonance~\cite{Wang19,Wang20}, can be incorporated to trapped-ion systems to achieve significantly  enhanced robustness in quantum information processing via spin-spin dynamics. In particular,  phase-adaptive DD methods use either random or correlated pulse phases to remove common experimental errors. We want to remark that, to vary the phase of each delivered pulse in a sequence represents a minimal added experimental cost while, in contrast, we show that it leads to a large improvement in the fidelity of the implemented quantum gates. We choose the system of MW-driven trapped ions in a static magnetic-field gradient to illustrate our proposal, although the results presented here can be incorporated to any setup presenting longitudinal qubit-boson coupling such as superconducting circuits~\cite{Manucharyan09,Richer16} or ions in Penning traps~\cite{Jain20}. In addition, we notice that in this work we consider MW Rabi frequencies orders of magnitude smaller than those in Ref.~\cite{Arrazola18}, making our proposal experimentally accesible for current setups, at the price of longer gate times.

The article is organised as follows: In Sec.~\ref{sec:system} we review how the longitudinal coupling in trapped ions leads to effective spin-spin interactions that can be exploited for quantum logic, or to implement quantum simulations. In Sec.~\ref{sec:sequence} we show that DD pulse sequences can be combined with these spin-spin interactions, achieving protection against environmental errors. We also find the consequences of using realistic finite-width MW pulses. In particular, we demonstrate that these pulses lead to a reduction of the effective spin-spin coupling strength. We will consider this effect in our newly designed sequences that use phase-adaptive methods. Finally, in Sec.~\ref{sec:Random} we incorporate phase-adaptive DD methods to generate robust spin-spin interactions in trapped-ion setups with longitudinal coupling. More specifically, via detailed numerical simulations we prove that phase-adaptive DD schemes offer a significant improved fidelity in quantum information processing with a minimal extra experimental cost (this is the appropriate control of each pulse phase). This demonstrates the usefulness of the presented method.

\section{Quantum logic with longitudinal coupling}\label{sec:system}

In this section, we briefly review the type of quantum dynamics one can get by exploiting trapped ions with longitudinal coupling. We consider a chain of $N$ trapped ions, placed in the axial ($z$) direction of a linear trap with an axial trapping frequency $\nu$, along a magnetic-field gradient $g_B$ in the same direction. The Hamiltonian that describes this situation is (here, and throughout the article, $H\rightarrow H/\hbar$ where $\hbar$ is the reduced Planck constant, meaning the Hamiltonians are given in units of angular frequency)
\begin{equation}\label{SysHamil}
H_{sys}=\sum_{j=1}^N\frac{\omega_{j}}{2}\sigma_j^z +\sum_{m=1}^{N}\nu_m a_m^\dagger a_m + 2\sum_{j,m}\eta_{j m}\nu_m(a_m+a_m^\dagger)|{\rm e}\rangle\langle {\rm e}|_j
\end{equation}
where $\sigma^z_j=|{\rm e}\rangle\langle{\rm e}|_j-|{\rm g}\rangle\langle{\rm g}|_j$, $\omega_j=\omega_0+\gamma_eB_j/2$ is the frequency of qubit $j$, related with the value of the magnetic field in the ion's equilibrium position $z_j^0$, i.e. $B_j=B_0+g_Bz_j^0$. Also, $\eta_{jm}=b_{jm}\frac{\gamma_eg_B}{4\nu_m}\sqrt{\frac{\hbar}{2M_I\nu_m}}$, $M_I$ is the mass of an ion, $\gamma_e$ is the electronic gyromagnetic ratio, $\nu_m$ and $a_m^\dagger (a_m)$ are the frequency and the creation (annihilation) operators associated with the $m$th normal vibrational mode, and $b_{jm}$ is a coefficient that relates this mode with the $j$th ion's displacement in the $z$ direction~\cite{James98}. 

The Schr\"odinger equation associated to Hamiltonian~(\ref{SysHamil}) can be analytically solved, and it describes an effective spin-spin interaction mediated by the vibrational modes~\cite{Mintert01,Wunderlich02,Porras04}. In order to see this, we first move to a rotating frame with respect to the first two terms in Eq.~(\ref{SysHamil}). The resulting Hamiltonian reads~\cite{Roos08}
\begin{equation}\label{SysHamilInt}
H_{sys}^I= \sum_{j,m}\eta_{j m}\nu_m(a_me^{-i\nu_m t}+a_m^\dagger e^{i\nu_m t})(1+\sigma_j^z).
\end{equation}
The time-evolution operator associated to Hamiltonian~(\ref{SysHamilInt}) has the form $U=U_SU_F$, where 
\begin{equation}\label{SpinForce}
U_F(t)=\prod_{j,m}^N\exp{[(\alpha_{jm}(t)a_m^\dagger-\alpha^*_{jm}(t)a_m)(1+\sigma_j^z)]}
\end{equation}
is a product of spin-dependent displacement operators with $\alpha_{jm}(t)=-\eta_{jm}(e^{i\nu_mt}-1)$, and 
\begin{equation}\label{SpinSpin}
U_S(t)=\exp{[-i\sum_{i,j}^N \theta_{ij}(t)(1+\sigma_i^z)(1+\sigma_j^z)]}
\end{equation}
is a spin-spin operator where $\theta_{ij}(t)=\sum_m\eta_{im}\eta_{jm}(\nu_mt-\sin{(\nu_mt)})$. If $\eta_{jm}\ll1$, the $U_F$ propagator can be ignored as $\alpha_{jm}(t)$ is a bounded quantity. Conversely, the quantity $\theta_{ij}(t)$ that appears in $U_S$ grows linearly with time $t$ leading to an effective spin-spin gate that is described by the Hamiltonian
\begin{equation}\label{EffHamil}
H^I_{\rm eff}=-\frac{\delta}{2}S^z -\sum_{i, j}^N J_{ij}\sigma_i^z\sigma_j^z
\end{equation}
where $\delta=\frac{\hbar\gamma_eg_B}{2M_I\nu}$, $S^\alpha=\sum_j^N\sigma_j^\alpha$, and $J_{ij}=\sum_m\nu_m \eta_{im}\eta_{jm}$. Notice that the effective Hamiltonian in Eq.~(\ref{EffHamil}) does not contain bosonic degrees of freedom.

Equation~(\ref{EffHamil}) corresponds to an Ising model with a longitudinal field~\cite{Ising25}. Interestingly, the dynamics of these type of Ising Hamiltonians can be combined with $\pi/2$ pulses and, with the help of  the Suzuki-Trotter expansion~\cite{Hatano05}, it leads to the simulation of distinct spin models~\cite{Zippilli14,Arrazola16}. In addition, the Ising model describes entangling operations between, e.g., two  ions in a chain~\cite{Piltz16}. To demonstrate this, we assume that all ions, excepting the $i$th and $j$th, are in the qubit's ground state $|\rm g\rangle$. In that case, Hamiltonian~(\ref{EffHamil}) reduces to
\begin{equation}\label{EffTwo}
H^I_{\rm eff}=-\frac{\delta_i}{2}\sigma_i^z -\frac{\delta_j}{2}\sigma_j^z -J\sigma_i^z\sigma_j^z,
\end{equation}
where $\delta_i=4\sum_{m=1}^N\nu_m\eta_{im}(\eta_{im}+\eta_{jm})$,  $\delta_j=4\sum_{m=1}^N\nu_m\eta_{jm}(\eta_{jm}+\eta_{im})$ (notice that a $J_0\sigma_j^z\sigma_k^z$ term contribute to the energy of the $j$th qubit with $-J_0$ if the $k$th qubit is projected in $|\rm g\rangle$ state) and $J=2\sum_{m=1}^N\nu_m\eta_{im}\eta_{jm}$. Now, by preparing the the initial state $|\!++\rangle$ ($\sigma^x|\pm\rangle=\pm|\pm\rangle$) for the $i$th and $j$th qubits and in an interaction picture with respect the first two terms in Eq.~(\ref{EffTwo}), the Bell state $|\!++\rangle+i|\!--\rangle$ (up to normalization) is obtained after a time $t_G=\pi/4J$. Regarding the fidelity of this procedure, in the absence of error sources this is only limited by the accuracy of the approximation $U_F\approx 1$. In particular, the Bell state fidelity can be bounded by
\begin{eqnarray}\label{FidelityApp}
F\approx 1-\frac{1}{2}\Big[\sum_{m,m'}\Theta_{mm'}(2\bar{n}_m+1)(2\bar{n}_{m'}+1)\Big],
\end{eqnarray}
where $\bar{n}_m$ is the average number of phonons in mode $m$ and $\Theta_{mm'}=|\alpha_{im}|^2|\alpha_{im'}|^2+|\alpha_{jm}|^2|\alpha_{jm'}|^2 + 4\alpha_{im}\alpha^*_{jm}\alpha_{im'}\alpha^*_{jm'}$ (see appendix A for further details on the calculation). Assuming that $\alpha_{jm}(t)=2\eta_{jm}$, which maximizes the value of $|\alpha_{jm}(t)|$, we obtain that, for $N=2$ and $\bar{n}_m\approx1$, 
\begin{equation}\label{FidelityBound}
1-F\lesssim 98\eta^4, 
\end{equation}
where $\eta=\frac{\gamma_eg_B}{4\nu}\sqrt{\frac{\hbar}{2M_{I}\nu}}$ is the effective LD factor (see appendix B for further details on the obtention of the bound). For two $^{171}$Yb$^+$ ions with a trapping frequency of $\nu\approx(2\pi)\times220$ kHz, the infidelity due to the residual spin-boson coupling is of the order of $10^{-3}$ for $g_B\approx150$ T/m ($\eta\approx0.056$) and of $10^{-5}$ for $g_B\approx50$ T/m ($\eta\approx0.018$). These conditions lead to gates times, $t_G$, of $t_G\approx270 \ \mu$s and $t_G\approx2.65$ ms, respectively.  Faster gates can be achieved by driving the system with continuous drivings~\cite{Bermudez12,Cohen15,Arrazola20} or with fast $\pi$ pulses~\cite{Arrazola18}. In these cases, $|\alpha_{jm}(t)|\leq2\eta_{jm}$ does not hold true (thus Eq.~(\ref{FidelityBound}) neither) however, Eq.~(\ref{FidelityApp}) is still valid to quantify the infidelity due to residual spin-boson coupling.

As it is shown in Ref.~\cite{Piltz13}, DD schemes are required to protect spin-spin dynamics against environmental noise. This can be achieved by applying sequences of $\pi$ pulses upon the qubits. However, these pulses also suffer from control errors (namely, deviations in their Rabi frequencies as well as detuning errors) whose effect on the spin-spin dynamics could be even more harmful than the noise introduced by the environment. Hence, DD sequences that are robust against these types of errors are highly desirable. These types of sequences have been considered in Ref.~\cite{Arrazola18} to generate fast entangling gates, with $\eta\gtrsim0.05$ and fast $\pi$ pulses i.e. with Rabi frequencies much larger than the trapping frequency. In that regime, the condition $U_F(t_G)=1$ is pursued by varying the time between pulses, and how to achieve this condition for more than two ions is still an unsolved problem~\cite{Arrazola18}. Here, we consider low-power MW pulses with Rabi frequencies below the trapping frequency and $\eta\approx0.01$. In this regime corresponding to current experimental setups~\cite{Piltz16,Weidt16} $U_F\approx 1$ is a good approximation, and, thus, the extension to more than two ions is straighforward. In the next section, we describe how DD sequences can be applied in this regime and analyse the effect of using low-power MW pulses for quantum information processing.

\section{Protected Multi-Qubit Dynamics}\label{sec:sequence}

\begin{figure}
\centering
\includegraphics[width=1\linewidth]{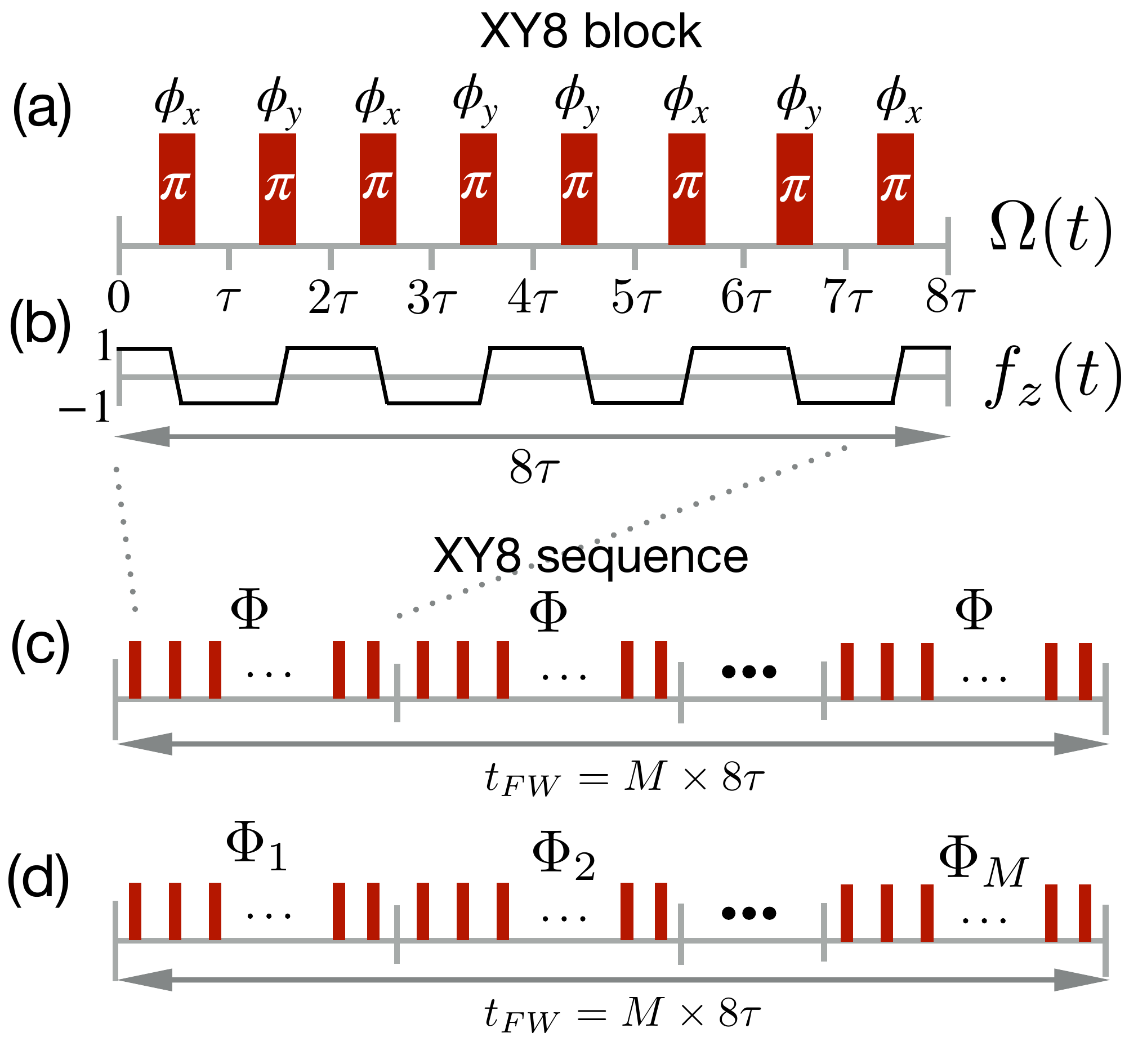}
\caption{ (a) XY8 pulse block with duration $8\tau$. The Rabi frequency takes the value $\Omega$ while applying the $\pi$ pulse, i.e. in the red region. The relative phase between the pulses changes according to $\phi_{x}=0+\Phi$ and $\phi_{y}=\pi/2+\Phi$. (b) Modulation function $f_z(t)$ corresponding to a single XY8 block. (c) The basic block shown in (a) is repeated to construct a longer sequence. In this case, the same global phase is chosen for all blocks. (d) The global phases $\Phi_{s}$ can be chosen to be different for each block $s$.}\label{fig:Pulses}
\end{figure}

For applying a pulsed DD sequence to the system described in the previous section, the ions have to be addressed with MW radiation. This requires a multi-tone MW signal that involves all qubit resonance frequencies $\omega_j$ and leading to the Hamiltonian 
\begin{equation}\label{SysPulse}
H=H_{sys}+\Omega(t)\sum_{j=1}^NS^x\cos{(\omega_j t-\phi)}.
\end{equation}
For simplicity, here we assumed that the Rabi frequency $\Omega(t)$ and the phase $\phi$ are the same for all ions. In an interaction picture with respect to $\sum_{j}\omega_j/2\sigma_j^z +\sum_{m}\nu_m a^\dagger_ma_m$, and neglecting terms rotating at frequencies $2\omega_j$ and $|\omega_j-\omega_i|$, we have
\begin{eqnarray}\label{SysPulse2}
H^I= H^I_{sys}
+\frac{\Omega(t)}{2}(S^+e^{i\phi} +S^-e^{-i\phi}).
\end{eqnarray}
In the case of pulsed DD techniques, MW radiation is delivered stroboscopically as $\pi$ pulses, such that $\Omega(t)$ is zero except when the pulse is applied, see Fig.~\ref{fig:Pulses}(a) for an specific example. As a consequence, the qubits get protected with respect to errors of the type $\sum_i \epsilon_i/2\sigma^z_i$ that may be caused  by, for example, variations on the magnetic field $B_0$. Contrary to Hamiltonian~(\ref{SysHamilInt}), the time-evolution operator associated to Eq.~(\ref{SysPulse2}) does not have an analytical form. In this manner, numerical simulations are required to study the performance of DD schemes to produce protected spin-spin dynamics.

In this work, we consider the XY8 pulse sequence, which is formed by blocks of $8$ pulses with an interpulse spacing $\tau$ and with phases $\vec{\phi}=\{0,\frac{\pi}{2},0,\frac{\pi}{2},\frac{\pi}{2},0,\frac{\pi}{2},0\}+\Phi$, where $\Phi$ is a global phase that adds to those of all pulses. In Fig.~\ref{fig:Pulses}(a), the specific case of an XY8 block is sketched. A pulse sequence may be formed by blocks with the same global phase $\Phi$, as in Fig.~\ref{fig:Pulses}(c), or different phases $\Phi_{s}$, as in Fig.~\ref{fig:Pulses}(d). In this section, we will consider the same phase $\Phi$ for all blocks, while in Sec.~\ref{sec:Random} we will prove the advantage of using different global phases for improved robustness against environmental and control errors. 

\begin{figure}
\centering
\includegraphics[width=1\linewidth]{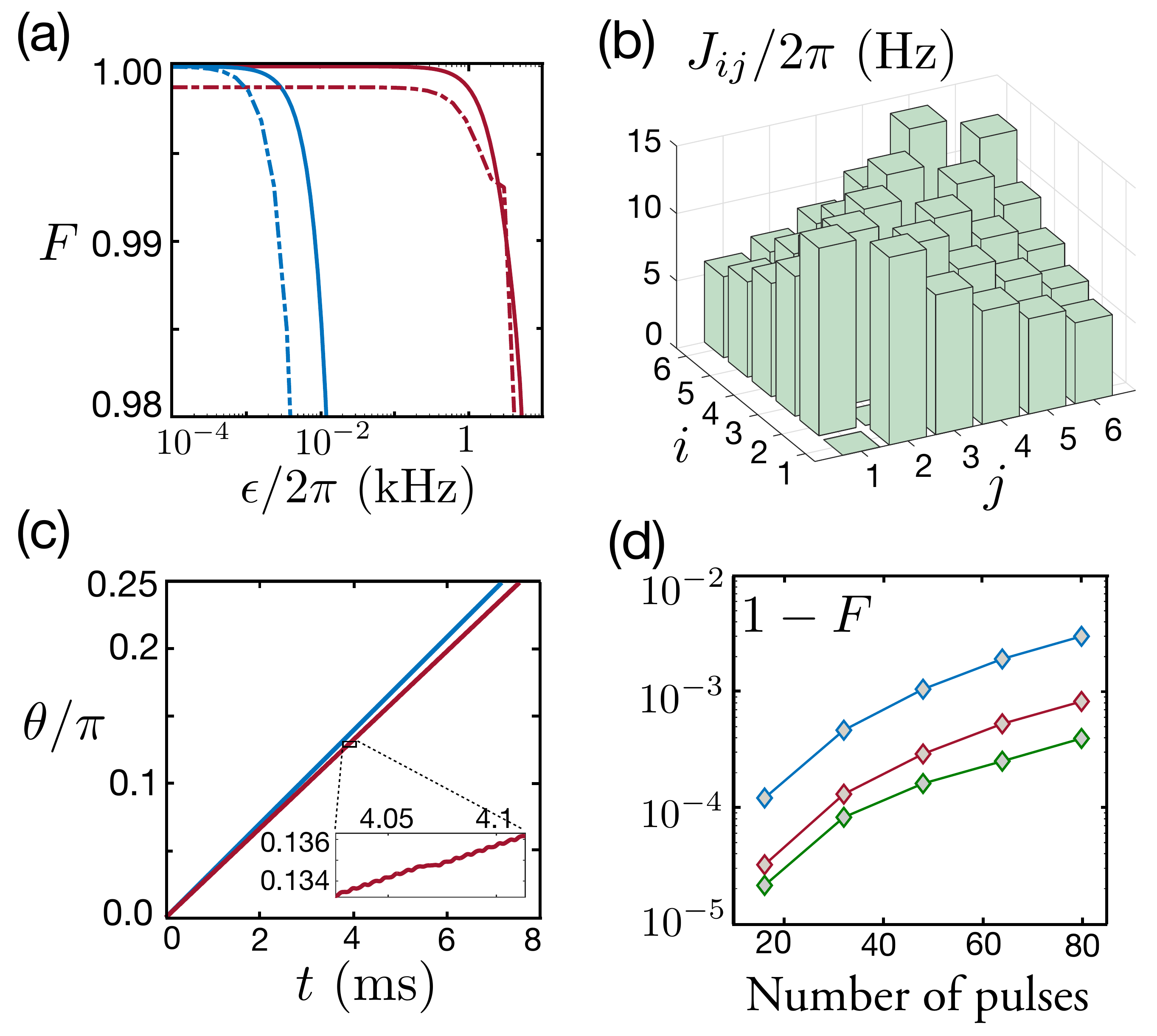}
\caption{ (a) Fidelity with respect the state generated by $H=\sum_{i,j}J_{ij}\sigma_i^z\sigma_j^z$ at $t_G=\pi/8\max{(|J_{ij}|)}$ on function of the strength of the unwanted shift $\epsilon$ and for $N=2$ (solid curves) and $N=6$ (dashed curves). The pulsed cases (red curves) are protected up to $\epsilon\approx(2\pi)\times1$ kHz, while the fidelity for the pulse-free cases (blue curves) takes values below $99\%$ around $\epsilon\approx(2\pi)\times5$ Hz; an improvement factor $\approx200$ is obtained in the pulsed case. Due to the effect of realistic finite-width pulses, the pulsed case for $N=6$ gets $\approx99.8\%$ fidelity for $\epsilon\rightarrow0$. (b) $J_{ij}$ matrix elements for $N=6, \eta=0.018$ and $\nu=(2\pi)\times220$ kHz. (c) Accumulated two-qubit phase versus time for  for $\eta=0.0113$, $\nu=(2\pi)\times200$ kHz and $\Omega=(2\pi)\times40$ kHz. The pulse-free case (blue) reaches $\theta=\pi/4$ around $t_G\approx 7.1$ ms, while for the pulsed case (red) it takes $t_{FW}\approx7.5$ ms. Inset: Phase accumulation around  $t\approx4.05-4.1$ ms. It can seen that phase accumulation stops during the application of a $\pi$ pulse leading to a plateau effect which lasts $\approx12.5 \ \mu$s. (d) Bell state infidelity for $\eta=0.0113$, $\nu=(2\pi)\times200$ kHz and for the XY8 pulse sequence with $16, 32, 48, 64$ and $80$ pulses. The cases with Rabi frequencies $\Omega=(2\pi) \times (20,40,$ and $60$) kHz are shown with blue, red and green colors, respectively. Due to the effect of transversal terms in Eq.~(\ref{SysPulse3}), the fidelity decreases when increasing the number of pulses or decreasing $\Omega$. }\label{fig:second}
\end{figure}

In Fig.~\ref{fig:second}(a) we plot the fidelity with respect to the state generated by $H=\sum_{i,j}J_{ij}\sigma_i^z\sigma_j^z$ at $t_G=\pi/8\max{(|J_{ij}|)}$ (for $N=2$, this is a Bell state) as a function of an unwanted energy shift $\epsilon$, and for a number of ions $N=2$ (solid) and $N=6$ (dashed). In particular, we will consider a scenario involving $\pi$ pulses, as well as the pulse-free case. The parameters we use in our simulations are $\nu=(2\pi)\times220$ kHz and $\eta=0.018$, for which the ideal gate times are $t_G=2.62$ ms and $t_G=4.50$ ms for $N=2$ and $N=6$, respectively. At this point we anticipate that, when combining the spin-spin Hamiltonian with finite-width $\pi$ pulses, a total time $t_G$ will no longer be sufficient in order to obtain the desired state. Furthermore, the corresponding values of $J_{ij}$ for $N=6$ are plotted in Fig.~\ref{fig:second}(b), while further details about their calculation are in Appendix~C. 

In the case of $N=2$, we compare the evolution produced by the ideal Hamiltonian $H=\sum_{i,j}J_{ij}\sigma_i^z\sigma_j^z$ and the Hamiltonian (\ref{SysPulse2}) that includes realistic finite-width pulses where, in addition, each bosonic mode is in a thermal state with $\bar{n}_m=1$. For $N=6$, due to the computational overhead of simulating 6 bosonic modes, we consider all motional modes to be in their respective ground states. In Fig.~\ref{fig:second}(a), the fidelity-curves in the pulse-free cases (solid-blue and dashed-blue curves) drop down to $99\%$ at $\epsilon\approx(2\pi)\times 5$ Hz. On the contrary, in the pulsed cases (solid-red and dashed-red curves), the fidelity falls to $99\%$ for $\epsilon\approx(2\pi)\times 1$~kHz, which represents an improvement factor of $\approx 200$. Further details regarding our simulations are: For the pulsed case with $N=2$, we considered a sequence with 64 pulses (this is 8 XY8 blocks) where each $\pi$ pulse has been implemented with a Rabi frequency  $\Omega=(2\pi)\times40$~kHz leading to a $\pi$ pulse time of $12.5 \ \mu$s. In addition, an interpulse spacing of $\tau=50.28 \ \mu$s have been employed. For $N=6$, the interpulse spacing is $\tau=146.56 \ \mu$s, and we considered a sequence of $32$ $\pi$ pulses (i.e. 4 XY8 blocks) with $\Omega=(2\pi)\times60$ kHz leading to a $\pi$ pulse time of $\approx 8.3 \ \mu$s. The final times for both sequences  that use finite-width pulses are, $t_{FW}=3.22$ ms ($64\times50.28 \ \mu$s) for the case with $N=2$, and $t_{FW}=4.69$ ms ($32\times146.56 \ \mu$s) for $N=6$. 

It is important to note that the latter final times $t_{FW}$ do not correspond to the ideal final times $t_G=2.62$~ms and $t_G=4.50$~ms. This is because, during the application of a $\pi$ pulse, the phase accumulation stops. Hence, for the pulsed cases, to achieve the same state as the one generated by the error-free Hamiltonian $H=\sum_{i,j}J_{ij}\sigma_i^z\sigma_j^z$ at $t_G$ requires a longer time $t_{FW}$. This effect will be later studied in more detail. Another aspect that can be observed in Fig.~\ref{fig:second}(a) is that, although the pulsed case is still far-less sensitive with respect the unwanted shifts than the pulsed-free case, the fidelity of the achieved state is, in the best case, around $99.8\%$ (note that the dashed-red curve does not reach the value $1$ when $\epsilon\rightarrow0$). Later, we will specify how this fidelity can be improved.

The previously commented effects, i.e. the reduction in the fidelity and the modified phase accumulation, are caused by the application of finite-length MW pulses. To show this, in Eq.~(\ref{SysPulse2}), we move to a rotating frame with respect to $\frac{\Omega(t)}{2}(S^+e^{i\phi} +S^-e^{-i\phi})$, which leads to the following Hamiltonian
\begin{equation}\label{SysPulse3}
\tilde{H}_{sys}^I= \sum_{j,m}\eta_{j m}\nu_m(a_me^{-i\nu_m t}+{\rm H.c.})(1+f_z(t)\sigma_j^z +f_\perp(t)\sigma_j^\perp).
\end{equation}
where $f_z(t)=\cos{(\int_{t_0}^{t}\Omega(t')dt' )}$, $f_\perp(t)=\sin{(\int_{t_0}^{t}\Omega(t')dt' )}$, and $\sigma_j^\perp=-i(\sigma_j^+e^{i\phi} -\sigma_j^-e^{-i\phi})$. During a $\pi$ pulse, the function $f_z(t)$ changes from $1$ to $-1$, see Fig.~\ref{fig:Pulses}(b), while $f_\perp(t)$ evolves from $0$ to $0$. Furthermore, when the system is not being driven, $f_z(t)$ is $1$ or $-1$ depending on the number of applied pulses, while $f_\perp(t)=0$. Now, for the sake of clarity in the presentation, we neglect the effect of the transversal terms -- i.e. those that appear multiplying $\sigma_j^\perp$ in Eq.~(\ref{SysPulse3}) -- as they are acting only during the $\pi$ pulse execution. However, they will be included in our numerical simulations, and their impact on the dynamics will be discussed. Finally, the spin-spin effective Hamiltonian takes the following form
\begin{equation}\label{EffHamil2}
\tilde{H}^I_{\rm eff}= -\sum_{i< j}^N J_{ij}(t)\sigma_i^z\sigma_j^z,
\end{equation}
where $ J_{ij}(t)=d\theta_{ij}(t)/dt$ with
\begin{equation}\label{phaseform}
\theta_{ij}(t)=\sum_m\nu_m^2 \eta_{im}\eta_{jm} \ \Im\int_{0}^{t}\int_{0}^{t'}e^{i\nu_m (t'-t'')} f_z(t')f_z(t'')dt' dt''.
\end{equation}
Note that, in the previous expression $\Im$ means the imaginary part of the subsequent integral. 

Most of the time, $f_z(t)$ takes the value $1$ or $-1$, resulting in a spin-spin coupling equivalent to the one in Eq.~(\ref{EffHamil}), i.e. $J_{ij}(t)=J_{ij}$ and the phase $\theta_{ij}(t)=J_{ij}\times t$. However, during the application of a $\pi$ pulse, $f_z(t)$ is not constant. The result is that the phase accumulated during $t_\pi$, i.e. the duration of the $\pi$ pulse, is smaller than $J_{ij}\times t_\pi$ (the latter is the phase that would be accumulated in during $t_\pi$ in case $f_z(t)$ would take a constant value). In practice, this implies that, when including realistic finite-width $\pi$ pulses, a longer time is needed to generate the final value for the phase than in the pulse-free case. This is shown in Fig.~\ref{fig:second}(c) for the case of two ions ($N=2$). In particular, we find that preparing a Bell state, this is to achieve the phase $\theta=\pi/4$, takes longer for the pulsed case  (solid-red curve) than for the pulse-free case in solid-blue. 
As a final comment, in Fig.~\ref{fig:second}(a) the appropriate final times, $t_{FW}$, for the pulsed cases have been calculated using Eq.~(\ref{phaseform}).

Regarding the transversal terms in Eq.~(\ref{SysPulse3})  we have neglected them in the presentation since they are non-zero only during the application of $\pi$ pulses. However, these transversal terms have a noticeable effect when the number of qubits is large, as well as with a growing number of applied MW pulses, and with significant pulse lengths (i.e. when the Rabi frequency of each pulse is small). In particular, these transversal terms are responsible for the $\approx99.8\%$ fidelity observed in Fig.~\ref{fig:second}(a) for the case with $N=6$. Here, the number of applied pulses is actually larger than in the $N=2$ case (note each ion is being addressed with a different MW pulse). This issue can be solved by introducing shorter $\pi$ pulses, i.e. with larger Rabi frequency. In appendix D, we show how the $\approx99.8\%$ fidelity can be improved up to $99.99\%$ by using a larger Rabi frequency $\Omega=(2\pi)\times160$ kHz and lowering the LD factor.

To further study the effect of these transversal terms, in Fig.~\ref{fig:second}(d) we plot the Bell state infidelity for sequences with different number of pulses and applied in two ions ($N=2$) without any unwanted shift $\epsilon$. More specifically, we compare XY8 sequences with Rabi frequencies $\Omega=(2\pi)\times 20$ kHz (blue curves), $\Omega=(2\pi)\times 40$ kHz (red curves), and $\Omega=(2\pi)\times 60$ kHz (green curves). In the simulations, we choose $\eta=0.0113$ ($g_B=26.8$ T/m) and $\nu=(2\pi)\times200$ kHz, and calculate the evolution according to Hamiltonian~(\ref{SysPulse2}) with each bosonic mode starting in a thermal state with $\bar{n}_m=0.5$. 

Using Eq.~(\ref{phaseform}), one can obtain the final time $t_{FW}$ that corresponds to a Bell state for the different Rabi frequencies. Notice that when decreasing the value of $\Omega$, the final fidelity decreases. We also notice that, for the Bell state preparation, introducing more pulses is detrimental due to the accumulated effect of the transversal terms. However, sequences with more pulses may offer an enhanced protection against environmental and control errors than their counterparts with reduced number of pulses. Hence, in a real experimental scenario, a trade-off between these two effects ought to be found, in order to achieve the best possible fidelities.

Finally, we consider the effect that the heating of the vibrational modes may have in the final fidelity of the prepared Bell states. For some gates this is a major source of infidelity, and specific techniques have been developed to gain robustness against this type of error~\cite{Haddadfarshi16,Webb18,Shapira18}. To account for the effect of motional heating we use a master equation of the form
\begin{equation}
\dot{\rho}=-i[H,\rho] + \sum_{m=1}^2\mathcal{L}_m(\rho)
\end{equation}
where $\rho$ is the density matrix, $H$ is Eq.~(\ref{SysPulse2}) in the Schr\"odinger picture, and the dissipative terms are
\begin{eqnarray}
\mathcal{L}_m(\rho)&=& \frac{\Gamma_m}{2}(\bar{N}_m+1)(2a_m\rho a_m^\dagger - a_m^\dagger a_m \rho - \rho a_m^\dagger a_m) \nonumber \\
&+& \frac{\Gamma_m}{2}\bar{N}_m(2 a_m^\dagger \rho a_m - a_m a_m^\dagger \rho - \rho a_m a_m^\dagger)
\end{eqnarray}
with $\bar{N}_m=[\exp{(\frac{\hbar\nu_m}{k_{\rm B}T})}-1]^{-1}$. Here, $\dot{\bar{n}}_m\approx\Gamma_m\bar{N}_m$ is the heating rate for the $m$th mode and $T=300$ K. For the parameters considered in Fig.~\ref{fig:second}(d), $\eta=0.0113$ and $\nu=(2\pi)\times200$ kHz, we consider heating rates of $\dot{\bar{n}}_1\approx107$ and $\dot{\bar{n}}_2\approx22.5$ phonons per second for the center-of-mass and breathing modes, respectively. These heating rates where derived using data from  Refs.~\cite{Weidt15,Weidt16} (for more details, check Appendix~E). Using the aforementioned heating rates, starting from the ground state of motion, and with $64$ pulses with $\Omega=(2\pi)\times60$ kHz and $t_{FW}=7.72$~ms, the infidelity increases from $\approx1.4\times10^{-4}$ to $\approx3\times10^{-4}$ when including heating in the model. This suggest that for Fig.~\ref{fig:second}(d) the heating of the motional modes may limit the infidelities to go below $\approx1.5\times10^{-4}$.

\section{Phase-Adaptive Pulse Sequences}\label{sec:Random}

\begin{figure*}[t]
\includegraphics[width=2\columnwidth]{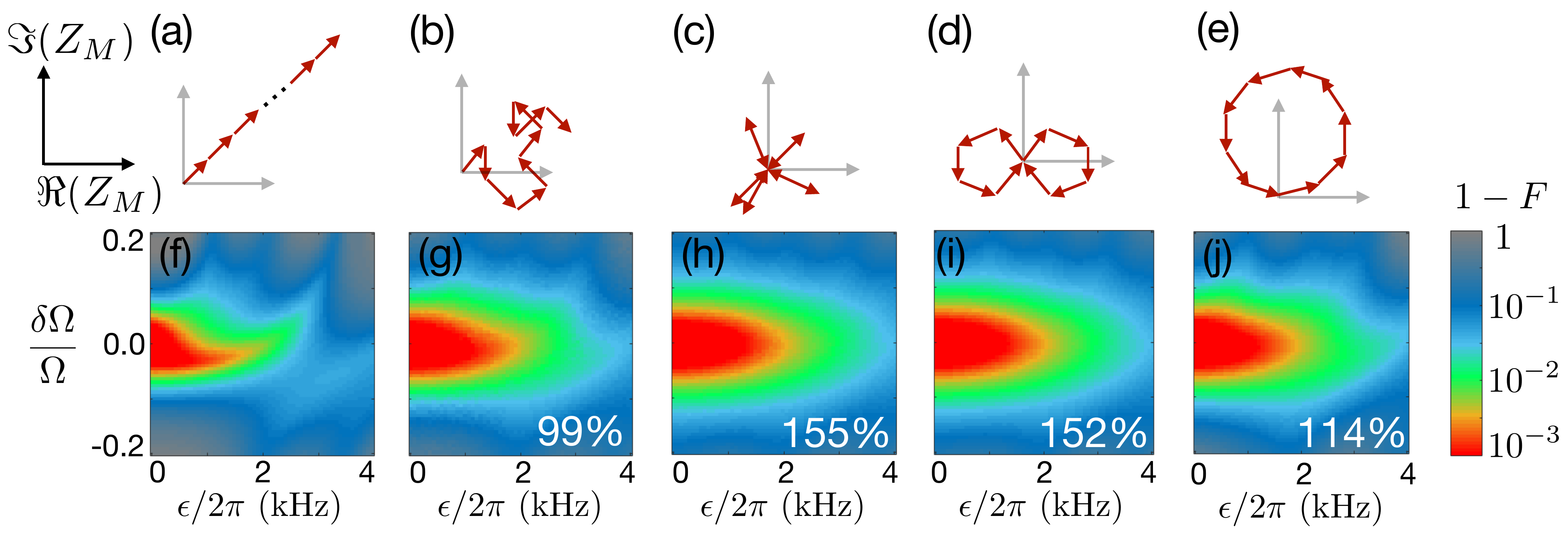}
\caption{XY8 pulse sequence with 80 pulses (i.e. 10 XY8 blocks) and different choices for the value of the phases $\Phi_{s}$. (a) Standard choice with $\Phi_{s}\equiv\Phi$. For each experimental run, $\Phi_{s}$ is chosen to be the same. (b) Full random phases (c), (d) and (e) Correlated phases where $\sum_{s=1}^S\exp{(-i\Phi_{s})}=0$ with $S=2,5$ and $10$ respectively. In all cases, the first global phase $\Phi_{1}$ is chosen randomly for each experimentla run. (f)-(j) Results of the described pulse sequences in terms the Bell state infidelity for different values of the static control error $\delta\Omega$ and qubit frequency shift $\epsilon$. The numbers in white represent how many more points with values above $99.9\%$ fidelity are with respect (f). For example, $100\%$ would mean there are twice as much points that fulfil this condition compared with (f).} 
\label{fig:RandomFig}
\end{figure*}

In this section, we show that an appropriate application of the global phase $\Phi$ on each DD block leads to  enhanced robustness of the quantum gates that can be engineered in trapped-ion systems presenting longitudinal coupling. This is the main result of our work. In particular, we implement two approaches for selecting the phase of each DD block. These are, firstly, a scenario where the phase $\Phi$ is randomly chosen in each block~\cite{Wang19} while, secondly, we  consider a situation where the phases of successive DD blocks are correlated in different manners~\cite{Wang20}. In our numerical simulations we will incorporate the corrected final time, $t_{FW}$, that appears as a consequence of using realistic finite-width pulses (notice we introduced $t_{FW}$ in the previous section). We anticipate that using the phase-adaptive method provides a significant enhancement in quantum information processing fidelities achieved by pulsed DD sequences, without adding a significant extra experimental complexity. Then our method is ready to be exploited in quantum platforms presenting longitudinal coupling, such as trapped ions and superconducting circuits.

Now, we present the basic mechanism leading to error cancelation by using the $\Phi$ phases. A single qubit under the effect of an imperfect MW field may be described by the following Hamiltonian:
\begin{equation}\label{ErrorHamiltonian}
H^I= \frac{\epsilon}{2}\sigma^z + \frac{\Omega+\delta\Omega}{2}(\sigma^+e^{i \phi} +\sigma^-e^{-i \phi}),
\end{equation}
where $\delta\Omega$ is constant shift in the Rabi frequency and $\epsilon$ accounts for a qubit frequency shift that may be produced by, for example, the variation of the intensity of the magnetic field $B_0$ or the effective coupling to other spins mediated by the collective motional modes. Like in the previous sections, in Eq.~(\ref{ErrorHamiltonian}) the bosonic degrees of freedom are omitted as $\eta_{jm}\ll1$. 

The application of an imperfect $\pi$ pulse is then characterised by the following matrix (see Appendix E for the derivation):
\begin{equation}\label{ImpMatrix}
U_\pi(\phi)=
\begin{bmatrix}
  \sin\gamma & ie^{i(\phi-\beta)}\cos{\gamma}\\
 ie^{-i(\phi-\beta)}\cos{\gamma} & \sin\gamma 
\end{bmatrix}.
\end{equation}
Here we have selected $\Phi =0$, while $\phi$ is the phase that determines the rotation axis -- e.g. $\phi = 0$ ($\pi/2$) means a $\pi$ pulse along the X (Y) axis  --  and $\beta$ and $\gamma$ are real numbers related with $\epsilon$ and $\delta\Omega$. When the deviations are zero, Eq.~(\ref{ImpMatrix}) corresponds to a perfect $\pi$ pulse. 

A general DD pulse block (with an even number of pulses) results in
\begin{equation}\label{BlockMatrix}
U_{\rm DD}=
\begin{bmatrix}
1 & iC\gamma\\
 iC^*\gamma &1 
\end{bmatrix}
+O(\gamma^2).
\end{equation}
Here, $C$ is a complex number that depends on the structure of the employed DD block, see Ref.~\cite{Wang19} for a detailed calculation. Repeating $M$ times this DD block leads to:
\begin{equation}\label{StandardMatrix}
U=
\begin{bmatrix}
1 & iMC\gamma\\
 iMC^*\gamma &1 
\end{bmatrix}
+O(\gamma^2).
\end{equation}
Changing the global phase of all pulses in the DD block as $\Phi \rightarrow\Phi+\xi $ would change $C$ in Eq.~(\ref{BlockMatrix}) to $Ce^{-i\xi}$. Then, if one choses different values for the phase $\Phi$ on each block (in the following we use $\Phi_{s}$ to denote the phase of the $s$th block) the matrix corresponding to an $M$-block sequence reads
\begin{equation}\label{RandomMatrix}
U=
\begin{bmatrix}
1 & iZ_{M}MC\gamma\\
 iZ^*_{M}MC^*\gamma &1 
\end{bmatrix}
+O(\gamma^2),
\end{equation}
where $Z_{M}=\frac{1}{M}\sum_{s=1}^M\exp{(-i\Phi_{s})}$. The off-diagonal elements in Eqs.~(\ref{StandardMatrix}) and (\ref{RandomMatrix}) are related with the robustness of the sequence in front of control errors. Essentially, if an even number of ideal $\pi$ pulses is applied (this is, $\pi$ pulses in absence of errors) the off-diagonal elements must be zero.  In Ref.~\cite{Wang19} the randomisation of the set of phases $\{\Phi_{s}\}$ was introduced to improve the performance of DD sequences for quantum sensing with NV centers.  More specifically, selecting random values for $\{\Phi_{s}\}$, induces a 2D random walk in $Z_{M}$ with a statistical variance $\langle |Z_{M}|^2 \rangle=1/M<1$. A further improvement for NMR detection purposes was introduced in Ref.~\cite{Wang20}. Here, the phases $\{\Phi_{s}\}$  are chosen such that $\sum_{s=1}^M\exp{(-i\Phi_{s})}=0$ (this is, the phases are correlated). In this case $\langle |Z_{M}|^2 \rangle=0$, thus the first-order dependence on the error parameter $\gamma$ gets removed.

Now, we show that these techniques (originally conceived for quantum detection of nuclear spins) are useful in the context of quantum information processing with trapped ions.  For that, we concatenate $10$ XY8 blocks and evaluate the robustness of a Bell state preparation protocol using different values for the phases $\Phi_{s}$. We use Hamiltonian~(\ref{SysPulse2}) as a starting point of our numerical simulations without doing any further approximation, while each bosonic mode starts at a thermal state with $\bar{n}_m=1$. More specifically, we study the robustness of the Bell state fidelity against errors in the energy of the qubits (or detuning errors) represented by $\epsilon$, and deviations $\delta \Omega$ in the delivered Rabi frequencies. 

The results are shown in Fig.~\ref{fig:RandomFig}. Note that we do not show the regions with negative $\epsilon$ as we find they are equivalent to those with positive $\epsilon$. In particular, in Fig.~\ref{fig:RandomFig}(a), we sketch the evolution of the quantity $Z_{M}$ in the complex plane, when all phases $\Phi_{s}$ take the same value. In Fig.~\ref{fig:RandomFig}(f), the Bell state infidelity is shown as a function of $\epsilon$ and $\delta \Omega$. The red region represent the highest Bell state fidelities that can reach values above $99.9\%$ for the following common experimental parameters: $\eta=0.0113, \nu=(2\pi)\times200$ kHz and $\Omega=(2\pi)\times40$ kHz. In Fig.~\ref{fig:RandomFig}(b), it is shown a possible behaviour for $Z_{M}$ when randomly selected phases $\Phi_{s}$ are employed. In this case, the robustness against control errors gets enhanced, as it can be seen in Fig.~\ref{fig:RandomFig}(g). In particular, in this Fig.~\ref{fig:RandomFig}(g) it can be observed a larger region that presents fidelities on Bell state preparation above $99.9\%$. More specifically, this region is $99\%$ larger than that in Fig.~\ref{fig:RandomFig}(f). The latter  corresponds to the application of the standard procedure where all phases $\Phi_{s}$ are equal. At this point we want to clarify that each point in Figs.~\ref{fig:RandomFig}(g)(h)(i) and (j) corresponds to the average $20$ different realisations, where in each realisation a different set of phases $\Phi_{s}$, random or correlated, has been employed.

In Figs.~\ref{fig:RandomFig}(c), (d) and (e), we show $Z_{M}$ for the cases where $\Phi_{s}$ are correlated such that $\sum_{s=1}^S\exp{(-i\Phi_{s})}=0$ with $S=2, 5$ and $10$, respectively. The performance of these sequences are shown in Figs.~\ref{fig:RandomFig}(h), (i) and (j). We find the largest fidelity enhancement with $S=2$. Comparing this case with the standard procedure where all phases $\Phi_{s}$ are equal, our phase-adaptive method presents an improvement of $155\%$. 

Phase-adaptive DD sequences can be directly applied to other schemes that make use of pulsed DD in trapped ions, for example, the one introduced in Ref.~\cite{Arrazola18}. The DD sequence considered in Ref.~\cite{Arrazola18} is the AXY-4 sequence, composed by four composite (each one formed by five $\pi$ pulses) pulses. As, in this case, each experimental run makes use of a single DD block, phase randomisation could be applied to all the different experimental runs. Then, by directly adding appropriate control on the phase on each DD block one can get significant improvements on quantum information processing fidelities.

\section{Conclusions}
We have introduced phase-adaptive pulsed DD methods in the context of trapped ions with longitudinal coupling. We showed that an optimal choice of the phases on each DD block leads to a significant enhancement in the robustness of entangling operations. This result has direct applications in quantum computing and simulation. Moreover, the enhancement achieved by our phase-adaptive method does not entail significant extra experimental cost, thus it can be easily incorporated to any pulsed DD method that aims to achieve robust quantum information processing with trapped ions or with other systems presenting longitudinal coupling such as superconducting circuits.

\section*{acknowledgements}
\begin{acknowledgements}
We acknowledge financial support from NSFC (11474193), SMSTC (2019SHZDZX01-ZX04, 18010500400 and 18ZR1415500), the Program for Eastern Scholar, Spanish Government via PGC2018-095113-B-I00 (MCIU/AEI/FEDER, UE),
Basque Government via IT986-16, as well as from QMiCS
(820505) and OpenSuperQ (820363) of the EU Flagship on Quantum Technologies, and the EU FET Open Grant Quromorphic (828826). I.~A.~acknowledges the UPV/EHU grant EHUrOPE.  X.~C.~acknowledges the Ram\'{o}n y Cajal program (RYC2017-22482) as well. This work was supported by Huawei HiQ funding for developing QAOA\&STA (Grant No. YBN2019115204). J.~C.~acknowledges the Ram\'{o}n y  Cajal program (RYC2018-025197-I)  and  the  EUR2020-112117 project of the Spanish MICINN, as well as supportfrom the UPV/EHU through the grant EHUrOPE.\end{acknowledgements}

\section*{Appendix A: Infidelity due to Residual Spin-Boson Coupling}
In the following, we consider how ignoring Eq.~(\ref{SpinForce})  affects the generation of a Bell state. The fidelity between two states $\rho_1$ and $\rho_2$ is given by
\begin{equation}\label{Fidelity1}
F=\frac{|{\rm Tr}(\rho_1 \rho_2^\dagger )|}{\sqrt{{\rm Tr}(\rho_1\rho_1^\dagger){\rm Tr}(\rho_2\rho_2^\dagger)}}.
\end{equation}
Thus, after a time $t=\pi/4J$, the Bell state fidelity is
\begin{equation}\label{Fidelity2}
F=\frac{\Big|{\rm Tr}\Big( \rho|B\rangle\langle B |\Big)\Big|}{\sqrt{{\rm Tr}(\rho\rho^\dagger) }},
\end{equation}
where $|B\rangle=\frac{1}{\sqrt{2}}(|++\rangle+i|--\rangle)$ with $|\pm\rangle_j=\frac{1}{\sqrt{2}} (|{\rm e}\rangle_j \pm |{\rm g}\rangle_j)$, and $\rho$ is the final state. The latter is given by
\begin{equation}\label{State1}
\rho={\rm Tr_M}\Big(U_F(t)(|B\rangle\langle B|\otimes\rho_T) U_F^\dagger(t)\Big),
\end{equation}
where $\rho_T$ is the state of the motional modes, and ${\rm Tr_M}$ stands for a partial trace of all $N$ motional subspaces. The matrix $\rho$ can be rewritten as 
\begin{equation}\label{DensityMatrixApp}
\rho=\frac{1}{4}
\begin{bmatrix}
1 & i\langle e^{g_2}\rangle & i\langle e^{g_1}\rangle & \langle e^{g_1+g_2}\rangle\\
-i\langle e^{-g_2}\rangle  & 1 & \langle e^{g_1-g_2} \rangle & -i\langle e^{g_1}\rangle\\
-i\langle e^{-g_1}\rangle  &  \langle e^{-g_1+g_2}\rangle  & 1 & -i\langle e^{g_2}\rangle\\
\langle e^{-g_1-g_2} \rangle &  i\langle e^{-g_1} \rangle & i\langle e^{-g_2} \rangle & 1
\end{bmatrix},
\end{equation}
up to a global phase, and where $g_j=2\sum_m\alpha_{jm}(t)a_{m}^\dagger -\alpha^*_{jm}(t)a_m$, $g_j^\dagger=-g_j$, $[g_i,g_j]=0$ and $\langle A\rangle={\rm Tr_M}\Big(A\rho_T\Big)$. Now, one can find that 
\begin{eqnarray}\label{Trace1}
{\rm Tr}\Big(\rho |B \rangle\langle B|\Big)=\frac{1}{4}\Big(1+\langle\cosh{(g_1)}\rangle+\langle\cosh{(g_2)}\rangle \nonumber\\+\langle\cosh{(g_1)}\cosh{(g_2)}\rangle\Big),
\end{eqnarray}
and 
\begin{eqnarray}\label{Trace2}
{\rm Tr}\Big(\rho\rho^\dagger \Big)=\frac{1}{4}\Big(1&+&\langle e^{g_1}\rangle\langle e^{-g_1}\rangle
+\langle e^{g_2}\rangle\langle e^{-g_2}\rangle \nonumber\\  
&+&\langle e^{g_1+g_2}\rangle\langle e^{-g_1-g_2}\rangle/2\nonumber \\
&+&\langle e^{g_1-g_2}\rangle\langle e^{-g_1 +g_2}\rangle/2  \Big).
\end{eqnarray}
If we assume that $|g_j|\ll1$, one can expand the exponential and the hyperbolic cosine functions up to the fifth power of $g_j$. If $\rho_T$ is a thermal state, the average value for odd powers of $g_j$ will be zero.  Using this, Eqs.~(\ref{Trace1}) and~(\ref{Trace2}) read 
\begin{eqnarray}\label{Trace3}
{\rm Tr}\Big(\rho |B \rangle\langle B|\Big)\approx 1+x +y +O(g^6)
\end{eqnarray}
and 
\begin{eqnarray}\label{Trace4}
{\rm Tr}\Big(\rho\rho^\dagger \Big)\approx 1+2x +z +O(g^6),
\end{eqnarray}
where
\begin{eqnarray}\label{xyz}
x&=&\frac{1}{4}(\langle g_i^2\rangle +\langle g_j^2\rangle) \nonumber\\
y&=&\frac{1}{3\cdot2^4}(\langle g_i^4\rangle +\langle g_j^4\rangle) +\frac{1}{2^4}\langle g_i^2g_j^2\rangle\\
z&=& \frac{1}{2^5}(2\langle g_i^2\rangle^2+2\langle g_j^2\rangle^2 + \langle (g_i+g_j)^2\rangle^2 +\langle (g_i-g_j)^2\rangle^2) \nonumber\\
&+&\frac{1}{3\cdot2^5}(2\langle g_i^4\rangle+2\langle g_j^4\rangle + \langle (g_i+g_j)^4\rangle +\langle (g_i-g_j)^4\rangle).\nonumber
\end{eqnarray}
The fidelity can then be expanded as
\begin{eqnarray}\label{Fidelity3}
F\approx (1+x+y)(1-x-z/2+3x^2/2),
\end{eqnarray}
which gives
\begin{eqnarray}\label{Fidelity4}
F\approx 1-\frac{1}{2^5}(\langle g_i^2\rangle^2+\langle g_j^2\rangle^2+4\langle g_ig_j\rangle^2).
\end{eqnarray}
Using that $\langle g^2_j\rangle=-4\sum_m|\alpha_{jm}(t)|^2(2\bar{n}_m+1)$ and $\langle g_ig_j\rangle=-4\sum_m\alpha_{im}(t)\alpha^*_{jm}(t)(2\bar{n}_m+1)$, Eq.~(\ref{Fidelity4}) can be rewritten as
\begin{eqnarray}\label{Fidelity6}
F\approx 1-\frac{1}{2}\Big[\sum_{m,m'}\Theta_{mm'}(2\bar{n}_m+1)(2\bar{n}_{m'}+1)\Big],
\end{eqnarray}
where
\begin{eqnarray}\label{BigOmega}
\Theta_{mm'}=|\alpha_{im}|^2|\alpha_{im'}|^2+|\alpha_{jm}|^2|\alpha_{jm'}|^2 + 4\alpha_{im}\alpha^*_{jm}\alpha_{im'}\alpha^*_{jm'}.
\end{eqnarray}

\section*{Appendix B: Fidelity Bound for a Two-Ion Crystal}

For $N=2$, and assuming $\bar{n}_m=1$, the fidelity formula in Eq.~(\ref{FidelityApp}) becomes
\begin{equation}
F\approx 1-\frac{9}{2}(\Theta_{11}+\Theta_{12}+\Theta_{21}+\Theta_{22}).
\end{equation}
If we assume that $\alpha_{jm}(t)=2|\eta_{jm}|$, which maximizes the value of $\alpha_{jm}(t)$, we have that
\begin{eqnarray}\label{fidelityIneq}
F\lesssim 1-2^3\cdot9(\eta_{11}^4+\eta_{21}^4+4\eta_{11}^2\eta_{21}^2\nonumber\\+2\eta_{11}^2\eta_{12}^2 +2\eta_{21}^2\eta_{22}^2+8\eta_{11}\eta_{12}\eta_{21}\eta_{22} \nonumber\\+\eta_{12}^4+\eta_{22}^4+4\eta_{12}^2\eta_{22}^2).
\end{eqnarray}
As $\eta_{im}=b_{im}(\nu_m/\nu)^{-3/2}\times \eta$, then $\eta_{11}=\eta_{21}=2^{-1/2}\eta$ and $\eta_{12}=-\eta_{22}=-2^{-1/2}\cdot3^{-3/4}\eta$. Using these expressions, Eq.~(\ref{fidelityIneq}) becomes
\begin{eqnarray}\label{fidelityIneq}
F\lesssim 1-2\cdot9(6-4\cdot3^{-3/2}+6 \cdot3^{-3})\eta^4,
\end{eqnarray}
which can be approximately rewritten as
\begin{eqnarray}\label{fidelityIneq}
F\lesssim 1-98\eta^4
\end{eqnarray}

\section*{Appendix C: Spin-Spin Coupling Matrix}

To characterize the ideal spin-spin Hamiltonian $H=\sum_{i,j}J_{ij}\sigma_i^z\sigma_j^z$ one has to calculate the spin-spin coupling matrix $J_{ij}$. This is given by $J_{ij}=\sum_m\nu_m \eta_{im}\eta_{jm}$, with $\eta_{im}=b_{im}(\nu_m/\nu)^{-3/2}\times \eta$. For $N=2$, $\{\nu_1,\nu_2\}=\{1,\sqrt{3}\}\times\nu$, and
\begin{eqnarray}
\vec{b}_{1}&=&\{0.7071,0.7071\} \nonumber\\
\vec{b}_{2}&=&\{-0.7071,0.7071\}
\end{eqnarray}
where $\vec{b}_{m}\equiv\{b_{1m},b_{2m}\}$. For $N=6$, $\{\nu_1,\nu_2,\nu_3,\nu_4,\nu_5,\nu_6\}=\{1,\sqrt{3},\sqrt{5.824},\sqrt{9.352},\sqrt{13.51},\sqrt{18.27}\}\times\nu$, and 
\begin{eqnarray}
\vec{b}_{1}&=&\{0.4082, 0.4082, 0.4082, 0.4082, 0.4082, 0.4082\}  \\
\vec{b}_{2}&=&\{-0.6080,-0.3433,-0.1118, 0.1118, 0.3433, 0.6080\}\nonumber \\
\vec{b}_{3}&=&\{-0.5531, 0.1332, 0.4199, 0.4199, 0.1332,-0.5531\} \nonumber\\
\vec{b}_{4}&=&\{0.3577,-0.5431,-0.2778, 0.2778, 0.5431,-0.3577\}\nonumber \\
\vec{b}_{5}&=&\{0.1655,-0.5618, 0.3963, 0.3963,-0.5618, 0.1655\} \nonumber\\
\vec{b}_{6}&=&\{-0.0490,0.2954,-0.6406, 0.6406,-0.2954,0.0490\},\nonumber
\end{eqnarray}
with $\vec{b}_{m}\equiv\{b_{1m},b_{2m},b_{3m},b_{4m},b_{5m},b_{6m}\}$~\cite{James98}.

\section*{Appendix D: Improved fidelity with a larger Rabi frequency}
Here, we use Hamiltonian (\ref{SysPulse2}) to demonstrate that improved fidelities are obtained when considering larger Rabi frequencies in the MW pulses. For that, we repeat the simulations done for Fig.~\ref{fig:second}(a) for the case $N=6$, however, in this case we consider $\Omega=(2\pi)\times160$ kHz. To avoid unwanted excitations of the motional modes, we increase the trapping frequency to $\nu=(2\pi)\times440$ kHz, and lower the effective LD factor to $\eta=0.008$. It can be observed that, unlike in Fig.~\ref{fig:second}(a), a fidelity close to $100\%$ (above $99.99\%$) is reached in the limit $\epsilon\rightarrow0$.
\begin{figure}
\centering
\includegraphics[width=0.8\linewidth]{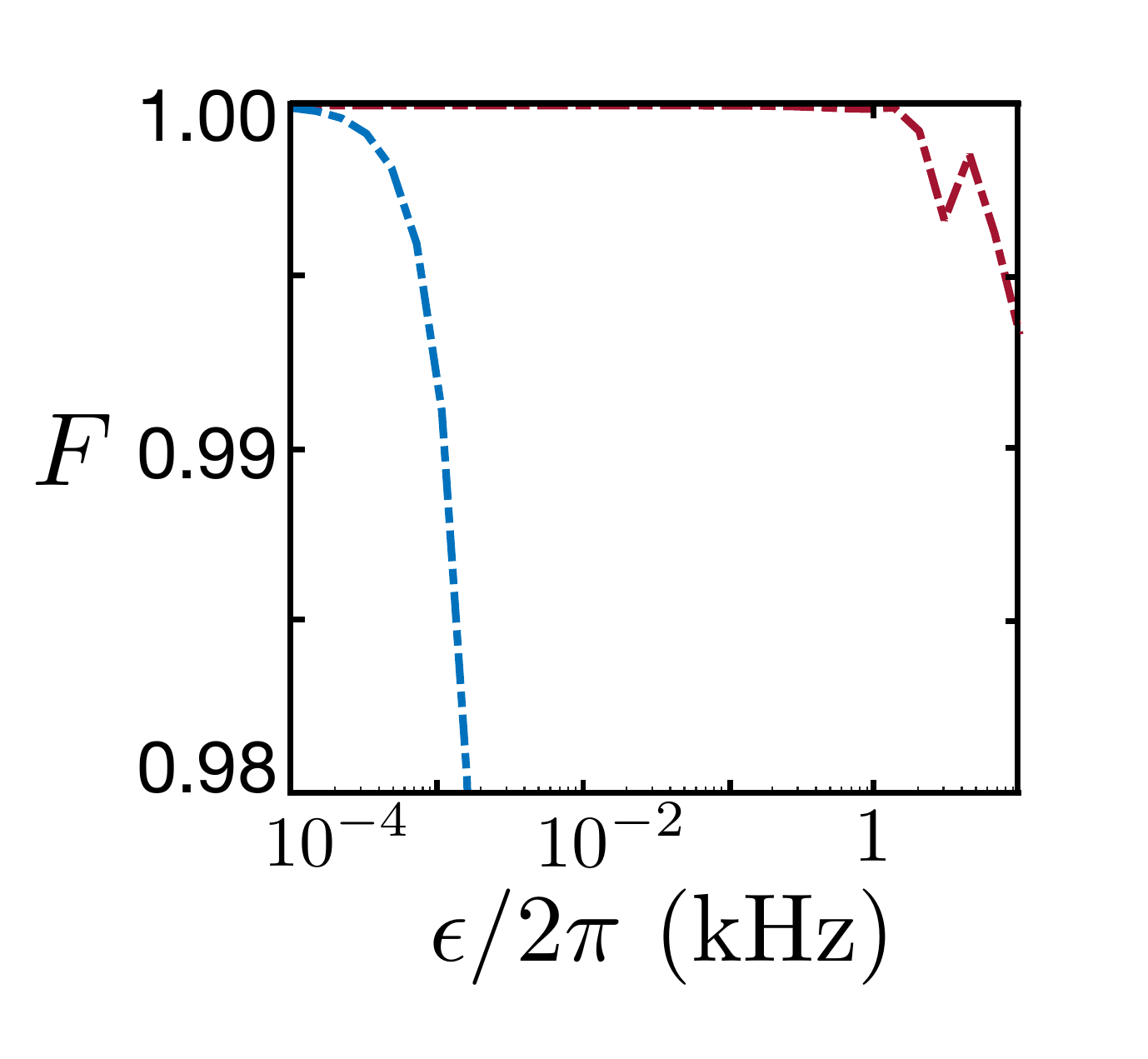}
\caption{Similar to Fig.~\ref{fig:second}(a), the fidelity with respect the state generated by $H=\sum_{i,j}J_{ij}\sigma_i^z\sigma_j^z$ at $t_G=\pi/8\max{(|J_{ij}|)}$ on function of the strength of the unwanted shift $\epsilon$ and for $N=6$ is shown. Here, for the pulsed case (dashed red) we consider a Rabi frequency of $\Omega=(2\pi)\times160$ kHz, which results in a better fidelity than the case with $\Omega=(2\pi)\times60$ kHz in Fig.~\ref{fig:second}(a) at the limit $\epsilon\rightarrow0$.}\label{fig:App}
\end{figure}

\section*{Appendix E: Derivation of the heating rates}

Here we explain how the heating rates used in the main text where derived. As reference values, we use the $\dot{\bar{n}}_{1,\rm ref}\approx41$ phonons/s reported for the center-of-mass mode with frequency $\nu_{1,\rm ref}=(2\pi)\times426.7$ kHz~\cite{Weidt15}, and the $\dot{\bar{n}}_{2,\rm ref}\approx1.3$ phonons/s assumed in Ref.~\cite{Weidt16} for the breathing mode with frequency $\nu_{2,\rm ref}=(2\pi)\times\sqrt{3}\times1.1$ MHz. In both cases, $g^{\rm ref}_B=23.6$ T/m and the ion-electrode distance is $d_{\rm ref}=310 \mu$m. To calculate the heating rates we use the following scaling relation~\cite{Brownnutt15},
\begin{equation}
\dot{\bar{n}}_m\approx \dot{\bar{n}}_{m,\rm ref} \Big(\frac{\nu_{m,\rm ref}}{\nu_m}\Big)^2 \Big(\frac{d_{\rm ref}}{d}\Big)^4,
\end{equation}
and assume that $g_B\propto 1/d^2$. 
\section*{Appendix F: Imperfect pulse unitary}
In an interaction picture with respect to $\frac{\Omega}{2}(\sigma^+e^{i\phi}+\sigma^-e^{-i\phi})$, Hamiltonian (\ref{ErrorHamiltonian}) reads
\begin{equation}
H^{II}(t)=\frac{\epsilon}{2}\sigma^ze^{-i\Omega(t-t_0)\sigma^\phi}+\frac{\delta\Omega}{2}\sigma^\phi,
\end{equation}
where $\sigma^\phi=\sigma^+e^{i\phi}+\sigma^-e^{-i\phi}$. If $t-t_0=\pi/\Omega$, then the unitary evolution corresponding to Hamiltonian (\ref{ErrorHamiltonian}) is
\begin{equation}\label{propagator}
U_\pi(\phi)=\exp{(i\pi/2\sigma^\phi)}\exp{[-i\frac{\pi \Gamma}{2\Omega}(\frac{\epsilon}{\Gamma}\sigma^z+\frac{\delta\Omega}{\Gamma}\sigma^\phi)]},
\end{equation}
where $\Gamma=\sqrt{\epsilon^2 +\delta\Omega^2}$. Using that, if $\hat{A}^2=1$, $e^{ia\hat{A}}=\cos{(a)}+i\sin{(a)}\hat{A}$, Eq.~(\ref{propagator}) can be rewritten as
\begin{equation}\label{PreImpMatrix}
U_\pi(\phi)=i\sigma^\phi
\begin{bmatrix}
\cos{\tilde{\Gamma}}-i\frac{\epsilon}{\Gamma}  \sin{\tilde{\Gamma}} & -i\frac{\delta\Omega}{\Gamma}\sin{\tilde{\Gamma}}e^{i\phi} \\
-i \frac{\delta\Omega}{\Gamma}\sin{\tilde{\Gamma}}e^{-i\phi} & \cos{\tilde{\Gamma}}+i\frac{\epsilon}{\Gamma}  \sin{\tilde{\Gamma}} 
\end{bmatrix},
\end{equation}
where $\tilde{\Gamma}=\frac{\pi\Gamma}{2\Omega}$. Eq.(\ref{PreImpMatrix}) can then simplified to 
\begin{equation}\label{ImpMatrix2}
U_\pi(\phi)=
\begin{bmatrix}
  \sin\gamma & ie^{i(\phi-\beta)}\cos{\gamma}\\
 ie^{-i(\phi-\beta)}\cos{\gamma} & \sin\gamma 
\end{bmatrix}, 
\end{equation}
where $\sin\gamma=\frac{\delta\Omega}{\Gamma}\sin{\tilde{\Gamma}}$ and $\tan\beta=\frac{\epsilon}{\Gamma}\tan{\tilde{\Gamma}}$.

\end{document}